\documentclass[a4paper,11pt]{article}
\usepackage{pos}
\usepackage{fontaxes}
\usepackage{mweights}
\begin{document}
\title{Clifford Fourier Transforms in (2+1)D Lattice Simulations of Soliton Propagations}

\author*[a]{Sadataka Furui}
\author[b]{Serge Dos Santos}

\affiliation[a]{Faculty of Science and Engineering, Teikyo University, \\Utsunomiya, 320 Japan }
\affiliation[b]{INSA Centre\, Val de Loire,  Blois,
Inserm U1253, Universit\'e de Tours, \\Imagerie et Cerveau, imaging and brain : iBrain, France}

\emailAdd{furui@umb.teikyo-u.ac.jp}
\emailAdd{serge.dossantos@insa-cvl.fr}
\abstract{Monte Carlo simulation of solitonic phonon propagation on a 2D plane in Weyl fermion-sea is analyzed.  We assume materials are filled with Weyl spinors located on $256\times 256$ 2D lattice points, which are expressed by quaternions ${\bf H}$.

The topology of solitonic phonon propagation is defined by modifying fixed point actions of 4D Quantum Chromo Dynamics action to (2+1)D action, replacing Dirac fermions by Weyl fermions, and changing the electric charge current flow to the energy flow. 

We consider $A$ type loops whose path are on a 2D plane, and $B$ type loops which contain two parallel links that connect two 2D plane on different time slices. The length of loops are restricted to be less than or equal to 8 lattice units. At the moment spatial lattice unit and time lattice unit are same. They can be chosen arbitrarily when one compares hysteresis effects with experimental data
 
Using the quaternion expression of Porteous, we calculate the plaquette part of loop actions and the link part of loop actions. Link actions of $A$ type loops cancel with each other, but those of $B$ type loops depends on whether the spin rotation is clockwise or that is counterclockwise. 

In the present work we consider average of clockwise rotating and counterclockwise rotating loop contributions. }

\FullConference{%
The 39th International Symposium on Lattice Field Theory,\\
8th-13th August, 2022,\\
Rheinische Friedrich-Wilhelms-Universität Bonn, Bonn, Germany
}
\maketitle

\section{Introduction}
In non-destructive-testing (NDT) the time reversal (TR) based nonlinear elastic wave spectroscopy (TR-NEWS) was successful. Use of time reversal mirrors (TRM) for enhancing signal to noise ratios of ultrasonic phonetic waves scattered in materials was proposed by Fink\cite{Fink97} and the technique of focussing, decomposition of time reversal operator (DORT) was developed\cite{RBCBF06}.

In TR-NEWS, transducers which emit phonons and their TR phonons and receivers are placed on sides of a 2 dimensional plane\cite{GDSBMC08,DSVP09}. Propagation of nonlinear phonetic waves in materials is recently reviewed in \cite{PM22}. 

Haldane\cite{Haldane04} revealed the spin quantum Hall effect (QHE) in space 2 dimensional TR violating electronic system can be explained through Chern index of algebraic topology. 
Kane and Mele\cite{KM05} studied the QHE in TR symmetric systems and showed that $Z_2$ topological order characterize presence of ordered phase on the Fermi surface.
Fidkowski and Kitaev\cite{FK11} discussed that TR invariant Majorana fermion spin 1 system has a trivial and the Haldane phase, and the breaking of $Z$ symmetry to $Z_2$ symmetry can be interpreted by real (R) edge spin states and quaternion (H) edge spin states. 
 Ryu et al.\cite{RML12} studied electromagnetic and gravitational responses and anomalies in topological insurators and superconductors by extending the theory of \cite{KM05}. They proposed anomaly cancellation between gravitation and electro-weak interaction. 

In 1982, Banks and Zaks \cite{BZ82} showed that in massless fermions systems, spontaneous chiral symmetry breaking does not occur.  
 The chiral symmetry breaking in Dirac fermion system was discussed in comparison with superconductivity in electronic media by Nambu and Jona-Lassinio\cite{NJL61}.
They pointed out without nonzero mass $m$, the particle would become an eigenstate of charge or chirality, and in the absence of Coulomb interactions the excitations become phonon-like. 

The convolution of a Khokhlov-Zaboltskaya solitonic wave obtained by Lapidus and Rudenko\cite{LR92}, and its TR solitonic wave show presence of anomalous zero mode\cite{SFDS21a}, and detailed analysis of TR-NEWS convolution data will present hints on the gravitational anomaly suggested in \cite{RML12}.

We consider a pair of Weyl spinor and its TR symmetric Weyl spinor in charge neutral media which violates spin rotational symmetry.
  When a soliton with effective mass is produced in fermion-sea, a zero-mode and energy gap in the infrared region could appear and the spontaneous chiral symmetry breaking can occur. If it does not originate from electro-weak interaction, the zero-mode need not be cancelled by gravitational anomaly. 
Since the media that a phonon propagates is not necessarily charged, we construct a model of Weyl fermions represented by quaternions sitting on $(2+1)D$ lattices\cite{SF21b,SF21c}. Instead of electric flow that propagates in the background of gravity, we consider the energy flow described by the DORT method.

To fix the path of phonons, we adopt the Fixed Point (FP) actions in the momentum space. They consist of 7 $A-$type loops on a 2D plane and 13 $B-$type loops that contain loops on two 2D planes on different time slices and two parallel links that are perpendicular to the 2D planes. We restrict ourselves to consider one loop graphs and total length of a loop is less than or equal to 8 lattice units. Hasenfratz and Niedermayer\cite{HN93} showed, in d=2 $O(3)$ non-linear $\sigma$ model, it is possible to simulate classical actions by discretized asymptotic free FP actions. 

At each cite a quaternion is placed and products of quaternions along the loop define the energy flow of the nonlinear waves. Since we consider situations in which phonetic solitons and TR solitons are emitted from the wall on the left hand side, the first step of the loop is chosen to be $e_1\Delta$, where $\Delta$ is the lattice unit  $\frac{1}{T_c N_t}$, where $T_c$ is the critical temperature, $N_t$ is the lattice size in the time direction,  $e_1$ is the unit vector along the $x_1$ axis. 
 Loops that start with $n e_1\Delta (n\geq 2)$ yield larger actions and we do not consider.
 The path of solitonic phonons are defined by minimizing the sum of 20 actions with optimum weight functions determined by the Monte Carlo (MC) simulation. The optimal superposition of TR-NEWS phonetic signals using memristor is demonstrated in \cite{DSHSF22} and the principle of reservoir computing adopted in memristors was reviewed in \cite{TYHNKTNNH19}.
The motivation of applying Clifford algebra to NDT is presented in \cite{SFDS22} and we present here details of lattice simulations and relations to the field theory.

\section{Weyl fermions and Clifford algebra}
The Weyl spinor and Majorana spinor in 4D are defined as 
\begin{equation}
\Psi\frac{1}{2}(1+\gamma_0{\bf u}), \quad \Psi\in Cl(1,3)
\end{equation}
where ${\bf u}=\pm \gamma_3$ for Weyl and ${\bf u}=\pm\gamma_1$ for Majorana. ($\gamma_0\gamma_1$ operator induces charge conjugations.)

 The Clifford Fourier Transform (CFT) in 2D is defined as 
\begin{equation}
{\mathcal F}\{f\}(\omega)=\int_{R^2}f({\bf x})e^{-i_2 \omega\cdot {\bf x}} d^2{\bf x}, \quad d^2{\bf x}=\frac{dx_1 \wedge dx_2}{ i_2},
\end{equation}
where $i_2$ is the unit bivector $e_1\wedge e_2$.

The nonlinear energy flow $f({\bf x})$ expressed by quaternions are Fourier transformed\cite{Hitzer22},
\begin{equation}
\hat f({\bf u})=\int_{R^2} e^{-{\bf i}x_1 u_1}f({\bf x})e^{-{\bf j}x_2 u_2}d^2{\bf x}.
\end{equation}

  The inverse Clifford transform (ICFT) of ${\mathcal F}\{f\}\in L^1(R^2;{\mathcal G}_2)$ is
 \begin{equation}
f({\bf x})=\frac{1}{(2\pi)^2}\int_{R^2} {\mathcal  F}({\bf \omega})e^{i_2 \omega}d^2{ \omega}\nonumber\\
  \end{equation}
  where  $\omega=u_1 e_1+u_2  e_2+\omega_{12}e_{12}$, ${\bf x}=x_1 e_1+x_2 e_2$,
 \begin{eqnarray}
 &&F=\{ f_1({\bf x,u}), f_2( {\bf x,u})\}, \nonumber\\
 &&f_1({\bf x, u})=2\pi e_1 x_1 u_1, f_2({\bf x, u})=2\pi e_2 x_2 u_2,
\end{eqnarray}
is supposed to be invertible. 

We let ${\bf u}$ be a $(2+1)D$ vector projected on $2D$ represented by quaternions, and transformation $\tau$ yields real vector space ${\bf x}=\tau({\bf u})$, where ${\bf u}\sim p_L({\bf u})$, where $L$ distinguishes 20 loops, that corresponds to 7 $A$ type loops and 13 $B$ type loops.
 
 The normalizing flow proposed by Papamakarios et al.\cite{PNRML21} for the convolutional neural network (CNN) can be applied to the search of 7 dimensional optimal weight functions of $A$ type actions.
 
 We transform the momentum space distribution $p_L({\bf u})=\sum_k^E p^k_L({\bf u})$ into random number distribution $(0,1)^E$, where $E$ is the number of epochs.  The distribution at the epoch $k$ is related to that at the epoch $k-1$ by ${\bf u}_k=T_k({\bf u}_{k-1})$. The inverse is ${\bf u}_{k-1}=T^{-1}_k({\bf u}_{k})$.
 
 To every distribution $u_i$, we consider the Clifford Fourier transform (CFT) $x_i=\tau(u_i, {\bf h}_i)$ where ${\bf h}_i=c_i({\bf u}_{<i})$, and $\tau$ is the transformer and $c_i$ is the conditioner at the $i$th epoch\cite{Hitzer22}.  
 
The position space transformation from $\bf u$ to $\bf x$, which is the inverse CFT is defined as
\begin{equation}
 h({\bf x})= {\mathcal F}^{-1}\{{\bf x}\}({\bf u})=\frac{1}{(2\pi)^2}\int_{R^2}e^{f x_1 u_1}{\mathcal F}\{ h\}({\bf u})
 e^{g x_2 u_2}d^2{\bf u}
\end{equation}
 where $d^2{\bf u}=d u_1 d u_2$, $f,g\in {\bf H}\sim Cl(0,2), f^2=g^2=-1$, or $f,g\in Cl(2,0)$. 
 
\section{Fixed point actions of one loop}
The plaquette actions and link actions of loops on Eucledian $(2+1)D$ space are mapped in quaternion space $\bf H$ as shown by Porteous\cite{Porteous95}. A unit quaternion $q\in \bf H$ satisfies $q^2=-1$ and for a vector $x$ there is anti-involution $x^-$ and a vector element is represented as
\begin{equation}
\left(\begin{array}{cc}
x & x x^-\\
1 & x^-\end{array}\right)=\left(\begin{array}{cc}
x & 0\\ 
0 & x^-\end{array}\right)
+\frac{1}{2}(1+x x^-)\left(\begin{array}{cc}
0 & 1\\
1 & 0\end{array}\right)+
\frac{1}{2}(1-x x^-)\left(\begin{array}{cc}
0 & -1\\
1 & 0\end{array}\right).
\end{equation}

 Its transformation can be expressed as
\begin{equation}
\left(\begin{array}{cc}
a&c\\
b&d\end{array}\right)\left(\begin{array}{cc}x & x x^-\\
1 & x^-\end{array}\right)
\left(\begin{array}{cc}
d^-&c^-\\
b^-&a^-\end{array}\right)=\lambda\left(\begin{array}{cc}
x'&x'x'^-\\
1& x'\end{array}\right).\\
\end{equation}
We interpret the real eigenvalue $\lambda=(bx+d)(bx+d)^-$ yields the plaquette action and $x x^-$ yields the link action.

 The matrix $\left(\begin{array}{cc}
a&c\\
b&d\end{array}\right)$ represents a special conformal transformation and is called Vahlen representation\cite{Vahlen02}. 

 In the case of $A$ type loops, we consider simple $2D$ QFT to obtain $f({\bf x})$, from ${\bf \omega}=\omega_{01}e_1+\omega_{02}e_2$ as in \cite{Hitzer22}. However, in the case of $B$ type loops, we  include the scalar term and consider the energy-time transformation.
We regard phonons as shock waves produced on Fermi surfaces and calculate plaquette actions derived from the fixed point action in $4D$\cite{DGHHN95} at $(2+1)D$ lattice points with the asymptotic high momentum action subtracted \cite{SF21b, SF21c, SFDS21a}. 
Link actions of $B$ type loops for large momenta are asymptotically zero, but they depend on the clockwise rotating (f-link) or counterclockwise rotating (e-link).  The relative strength of actions of e-link and f-link type depends on the random number sets used in MC. The average of e-link action and f-link action which are negative in the infrared region, plus plaquette action which are similar to that of $A$ type, are used. A solution of the traveling salesman problem\cite{PM96} says that, $A-$ type loops contain altogether 13 orderings of 7 weights, and the $B-$ type loops contain altogether 49 orderings of 13 weights.
 
In the case of $256\times 16$ lattices, the maximum and minimum of the total action of $u_2/\Delta=0$ was about twice of those of $u_2/\Delta=1$. 
This property reflects properties of CFT for $d=2,3$(mod 4) remarked by Hitzer\cite{Hitzer22}.
For a multi-vector $A_r\in{\mathcal G}_d$ of odd grade  $r=2s+1$ or even grade $r=2s$
\begin{equation}
A_{2s+1}i_d=-i_d A_{2s+1}, \quad A_{2s}i_d=+i_d A_{2s}.
\end{equation}

The qualitative difference of e-link actions and f-link actions does not change by the lattice size of $u_2$ direction.
The grade dependence observed in $256\times 16$ lattices was, however due to the asymmetry of $u_1$ vs $u_2$. We checked the grade dependence by calculating the convolution of the plaquette part and the link part of
the spinor and its TR spinor for each loop on $256\times 256$ lattices by using Mathematica\cite{Mathematica12}, and calculated their Fast Fourier Transform by using the the Machine Learning (ML) module torchfft\cite{RLM22}.

\section{Convolution of Weyl spinor and its TR spinor}
 The convolution of $a,b\in L^1(R^{2,0}; Cl(2,0))$ is
\begin{equation}
(a*b)({\bf x})=\int_{R^{2,0}}a({\bf y})b({\bf x}-{\bf y}) d^2{\bf y}
\end{equation}

We calculated plaquette actions and link actions of 20 FP actions of $256\times 256$ lattices.
The plaquette part of the convolution of the action and its TR action of $L1, L2$ and $L18$ of $x_1/\Delta\leq 128$ are presented in Fig.1. The loops projected on the 2D planes are shown in Fig.2.  The loop $L18$ starts from the middle of the lower edge, so the link of $2\Delta$ exists in the $u_2$ direction. The scale in the $(u_1,u_2)$ plane of $L18$ plaquette is twice of that of $L1$, however the convolution in the $(x_1,x_2)$ plane at largest $x/\Delta$ of $L18$ , $L1$ and $L2$ are almost the same, due to our specific renormalization.

 In most $B-$type loops f-link dominates over e-link. However, the convolution of the link and its TR link of the $L8$ which contains two parallel links that connect the upper 2D plane and the lower 2D plane with distance $\Delta$, as shown in Fig.4 has relatively large e-link as compared to the $L3$, as shown in Fig.3, 

For detecting anomalous scattering position, we make a superposition of the inverse Clifford transformed loop wave functions with the optimum weight. A solution of the traveling salesman problem says that, $A-$ type loops contain altogether 13 orderings of 7 weights, and the $B-$ type loops contain altogether 49 orderings of 13 weights. Fig.5 shows average of $A$ type actions which are almost loop independent, and $B$ type total actions of $L8$ (upper boundary) and $L10$  (lower boundary) and the average. 

The loop dependence of $B$ type total actions at $u_2=0$ as a function of $u_1$ is shown in Fig.6.   The line $p0_8$ and $p0_{10}$ mean $L8$ and $L10$ at $u_2=0$. 
There are pairs which have the overlapping path on a projected 2D plane but position of the red circle or the blue circle is shifted, like $L4-L26$, $L8-L10$, $L13-L15$ and $L16-L17$. Except $L8-L10$, there appear a point where the action at $(u_1,u_2)$ coincide as shown in Fig.7. We expect only the $L8-L10$ action pair is coherent, and other pairs may not play important roles in detection of hysteresis effects.

The action is large when the distance between the red circle and the blue circle is large. When $x_2=0$, the convolution of each total action and TR total action has a peak at around $x_2/\Delta\sim 22$.  
The dependence of weight function of the fixed point $L8$ total action given by random number samples is not significant as shown in Fig.8.

\begin{figure*}[h]
\begin{minipage}{0.47\linewidth}
\begin{center}
\includegraphics[width=5.cm,angle=0,clip]{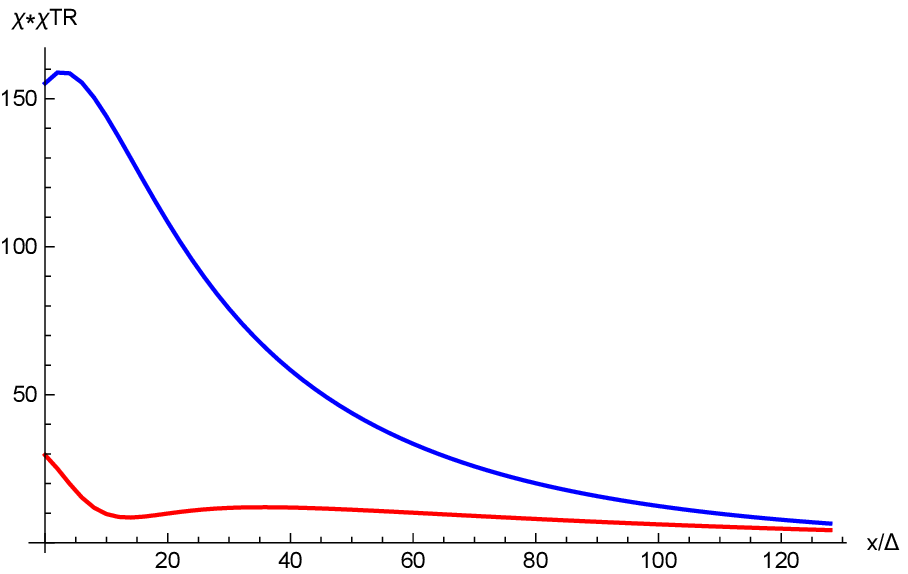}
\end{center}
\end{minipage}
\begin{minipage}{0.47\linewidth}
\begin{center}
\includegraphics[width=5.cm,angle=0,clip]{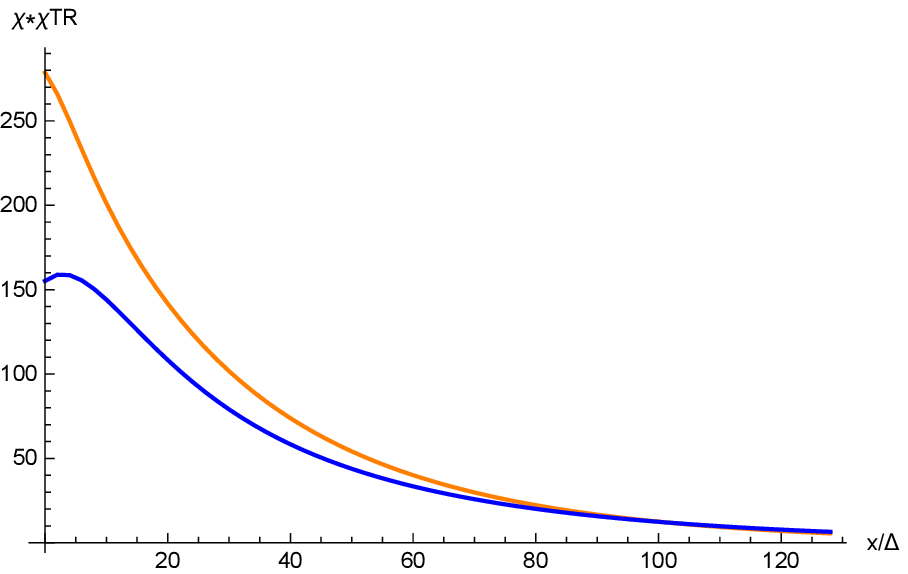}
\end{center}
\end{minipage}
\caption{ The convolution of the plaquette action and its TR action of $L1$ (blue line) v.s. $L18$ (red line) (left)  and those of $L1$ (blue line) v.s. $L2$ (red line) as a function of $x_1/\Delta$ at $x_2=0$ (left).  .}
\begin{minipage}[h]{0.47\linewidth}
\begin{center}
\includegraphics[width=3cm,angle=0,clip]{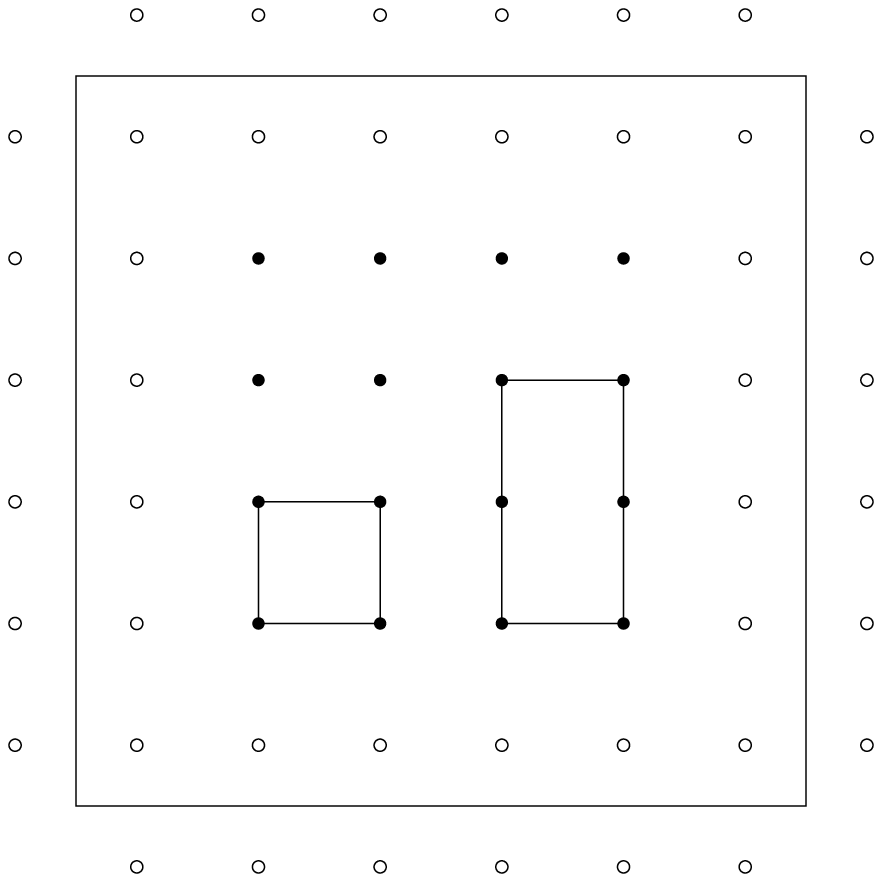} 
\end{center}
\end{minipage}
\hfill
\begin{minipage}[h]{0.47\linewidth}
\begin{center}
\includegraphics[width=3cm,angle=0,clip]{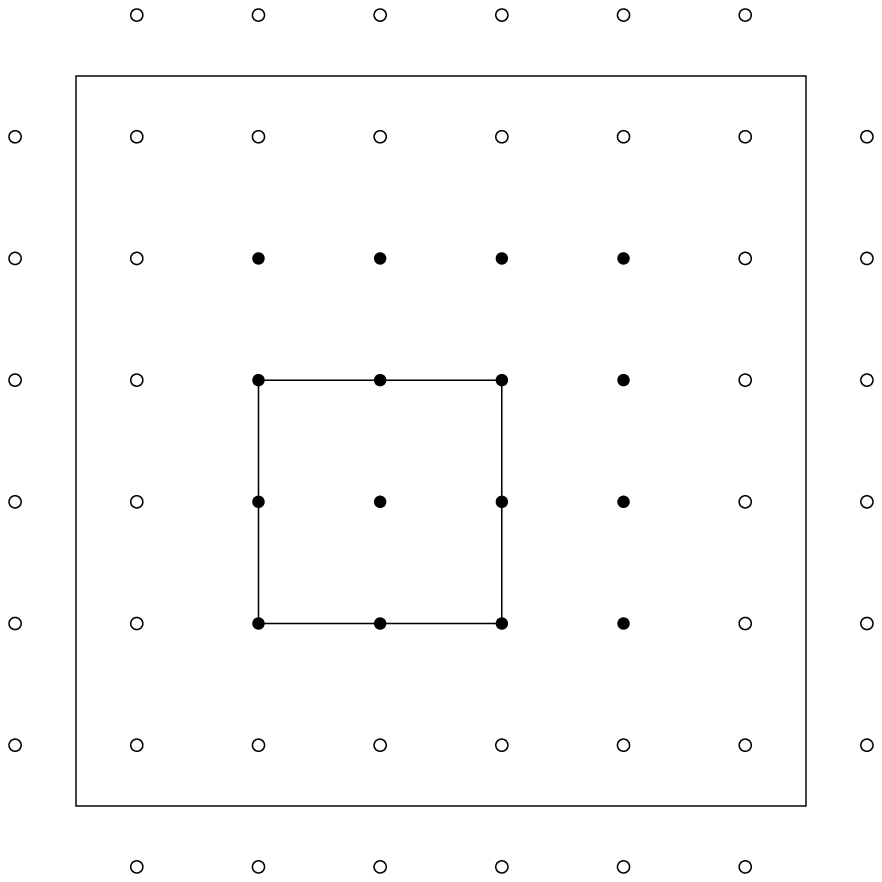} 
\end{center}
\end{minipage}
\caption{The $L1$ and the $L2$ loop (left) and the $L18$ loop (right) on a 2D lattice. .}
\begin{minipage}{0.47\linewidth}
\begin{center}
\includegraphics[width=5.cm,angle=0,clip]{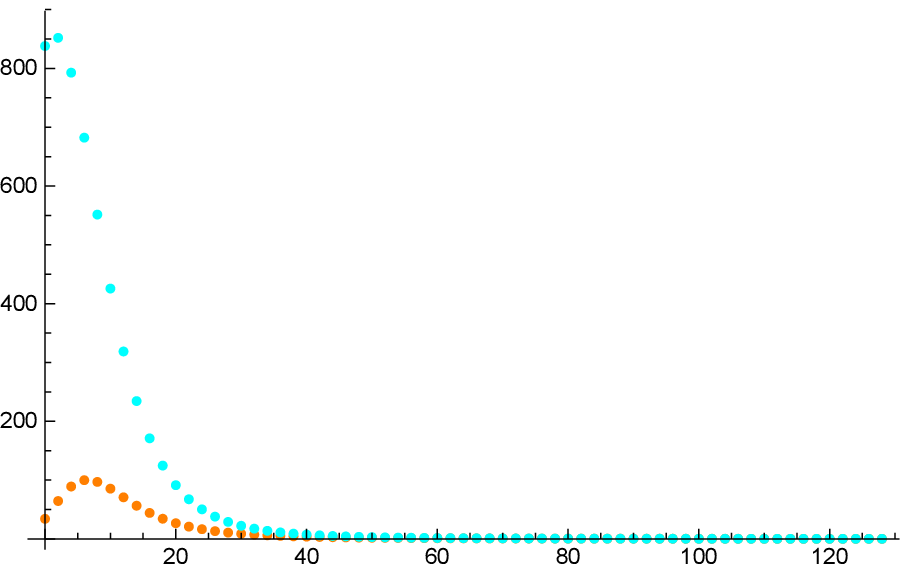}
\end{center}
\end{minipage}
\hfill
\begin{minipage}{0.47\linewidth}
\begin{center}
\includegraphics[width=5.cm,angle=0,clip]{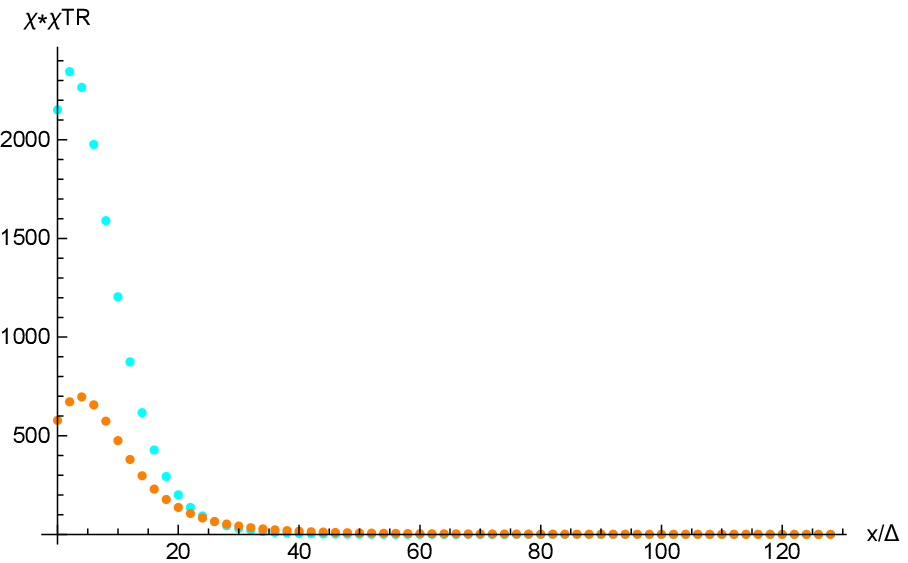}
\end{center}
\end{minipage}
\caption{ The convolution of the e-link action and its TR action (orange) and that of the f-link action and its TR action (paleblue) of the $L3$ loop (left) and those of the $L8$ loop (right).  }
\begin{minipage}[h]{0.47\linewidth}
\begin{center}
\includegraphics[width=3cm,angle=0,clip]{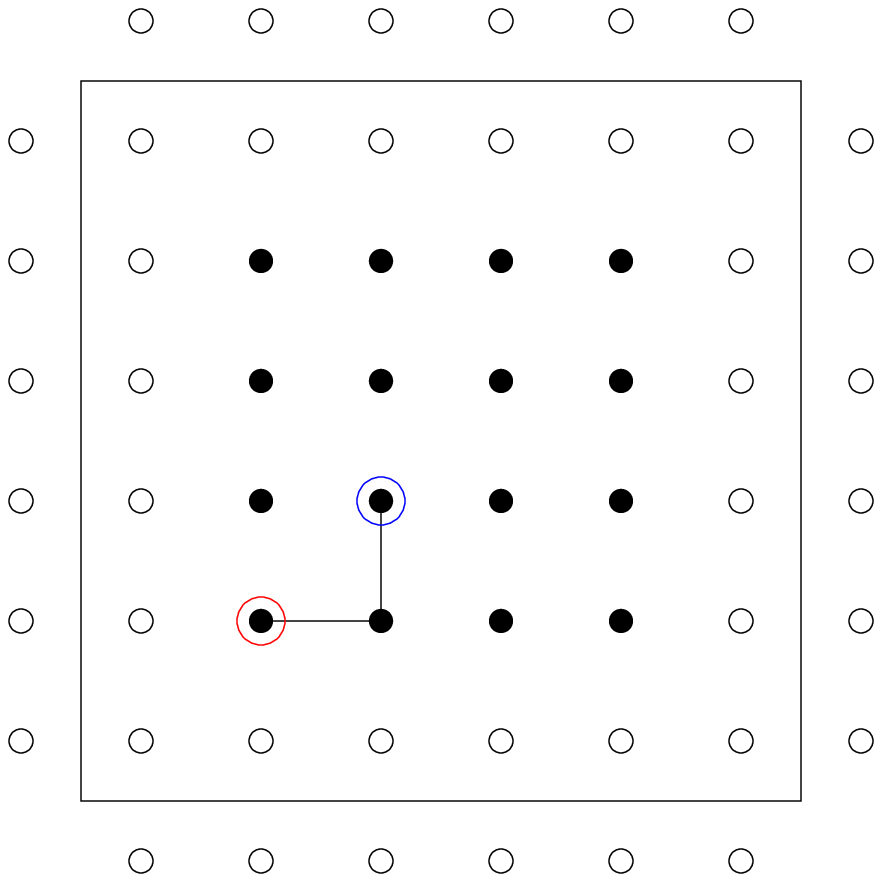} 
\end{center}
\end{minipage}
\hfill
\begin{minipage}[h]{0.47\linewidth}
\begin{center}
\includegraphics[width=3cm,angle=0,clip]{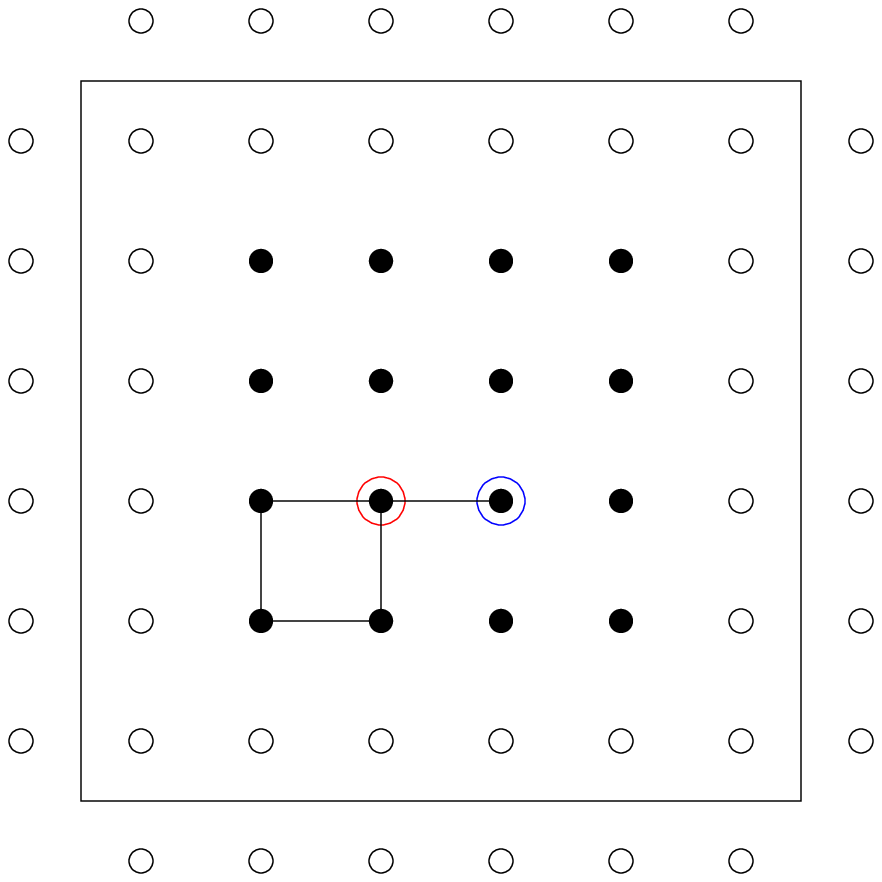} 
\end{center}
\end{minipage}
\caption{The $L3$ loop (left) and the $L8$ loop (right) on a 2D lattice. The red circle of $L8$ shifted to the right by $\Delta$ yields the $L10$ loop}
\end{figure*}
\begin{figure*}[htb]
\begin{minipage}[h]{0.47\linewidth}
\begin{center}
\includegraphics[width=5.5cm,angle=0,clip]{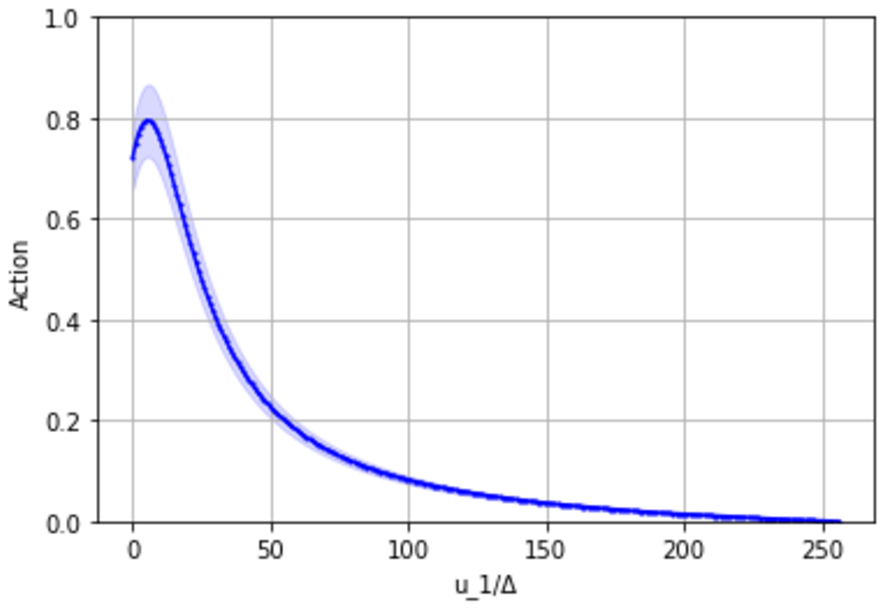}
\end{center}
\end{minipage}
\hfill
\begin{minipage}[h]{0.47\linewidth}
\begin{center}
\includegraphics[width=5.5cm,angle=0,clip]{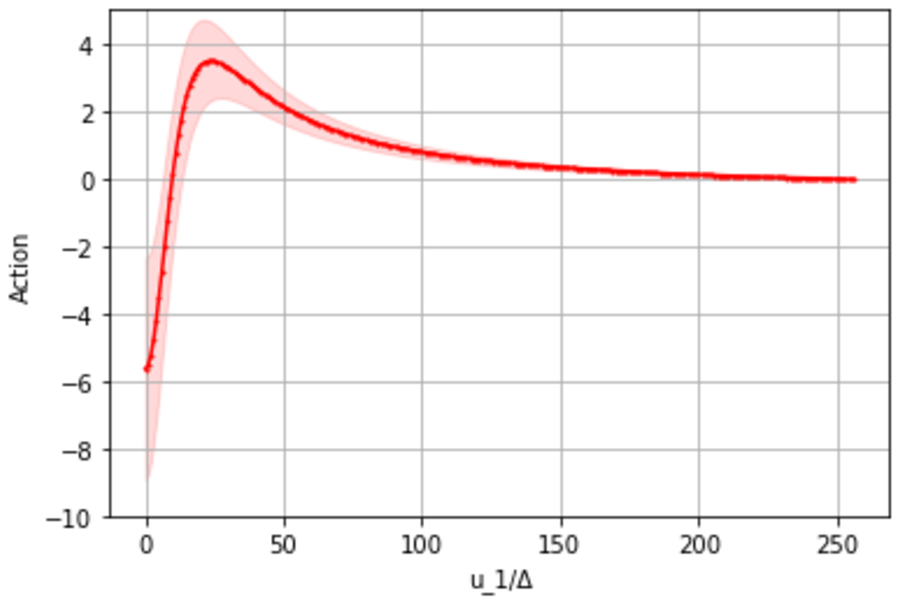}
\end{center}
\end{minipage}
\caption{The sum of $A-$type plaquette actions (left) and $B-$type plaquette+link actions of $L8$, $L10$ and the average (right) in momentum space $u_1/\Delta=[0,256]$, $u_2=0$.  }
\begin{minipage}[h]{0.47\linewidth}
\begin{center}
\includegraphics[width=5.5cm,angle=0,clip]{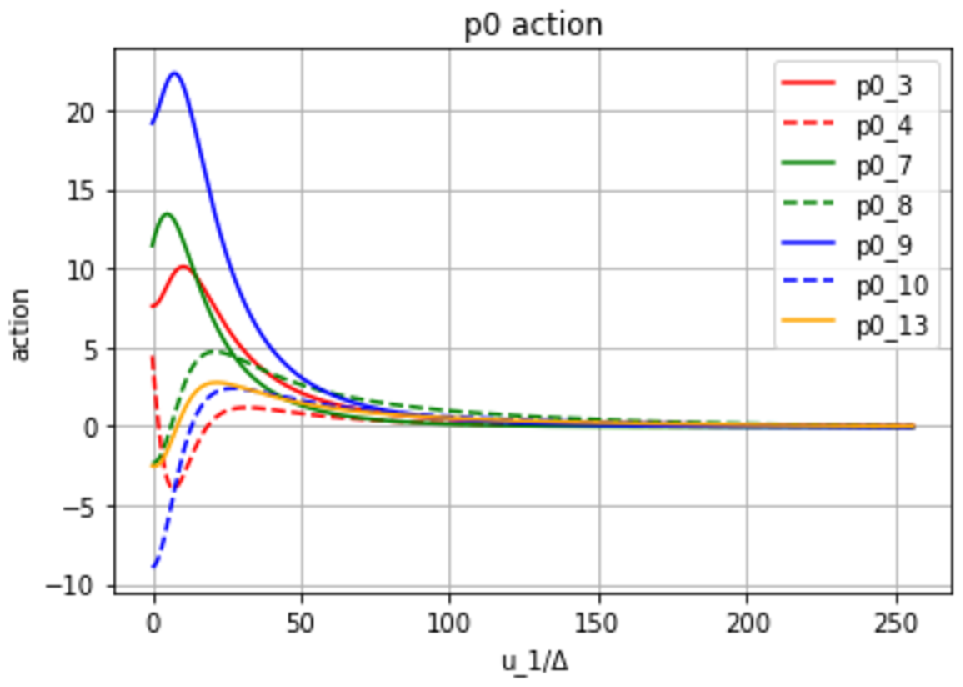}
\end{center}
\end{minipage}
\hfill
\begin{minipage}[h]{0.47\linewidth}
\begin{center}
\includegraphics[width=5.5cm,angle=0,clip]{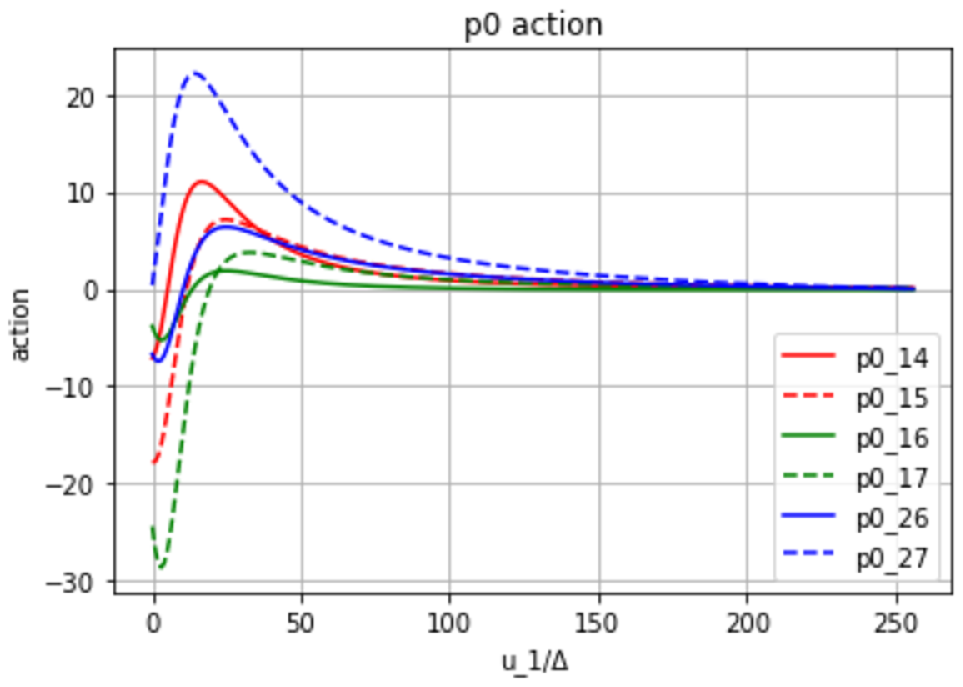} 
\end{center}
\end{minipage}
\caption{The loop dependence of $B-$type total action of $L3-L13$ (left) and $L14-L27$ (right) in momentum space $u_1/\Delta=[0,256]$, $u_2=0$. }
\begin{minipage}[h]{0.47\linewidth}
\begin{center}
\includegraphics[width=5.5cm,angle=0,clip]{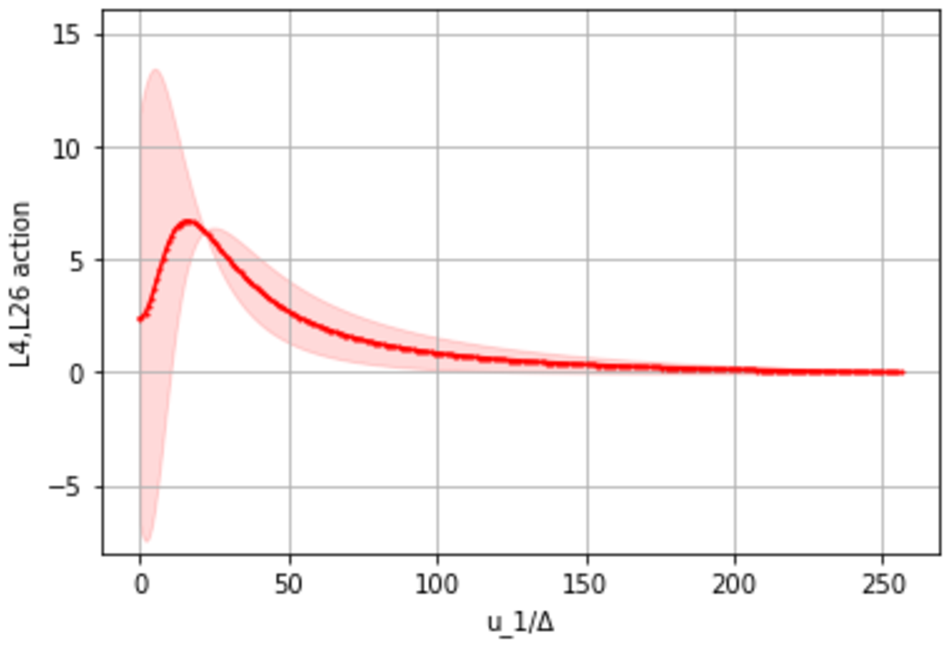}
\end{center}
\caption{The sum of plaquette+link actions of $L4$, $L26$ and the average in momentum space at $u_2=0$. }
\end{minipage}
\hfill
\begin{minipage}[h]{0.47\linewidth}
\begin{center}
\includegraphics[width=5.5cm,angle=0,clip]{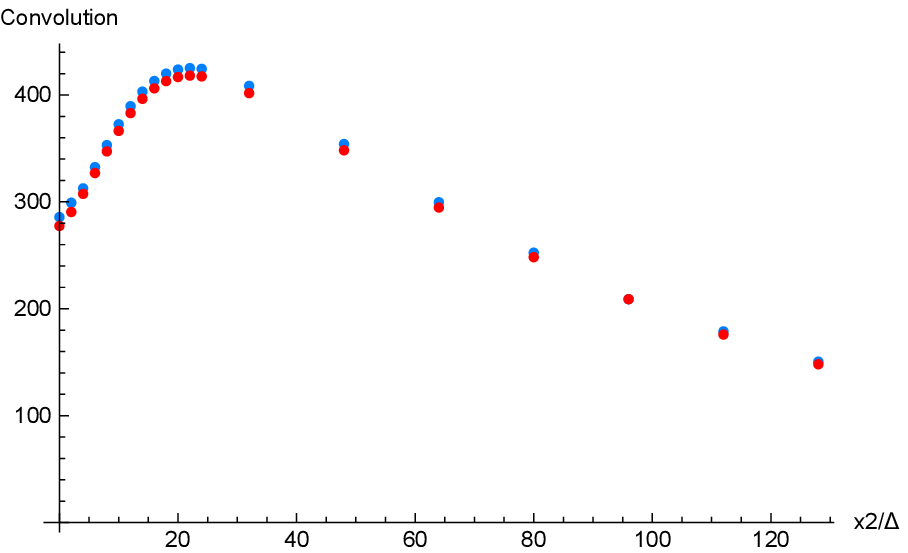}
\end{center}
\caption{The weight function dependence of the convolution of $L8$ . $x_2=0$, $0\leq x_1/\Delta\leq 128$. Red points are 'optimum' $\gamma$ sampe and blue points are a non-optimum sample.}
\end{minipage}
\end{figure*}

\section{Conclusion and discussion}
 For optimization of the weight function in ML, we first apply the CNN using the gradient decent method\cite{RLM22, Friedman01}.  We take the parameter space $(y,{\bf x})$, where $y$ specifies the random number set and ${\bf x}$ specifies the 20 loops. Instead of shuffling in ML, we choose paths given by the solution of traveling salesman problem and train the system to find the minimum action as a combination of the 20 FP actions. Actions of each loop are calculated by Mathematica and stored in the workstation

One tries to minimize the loss function $L(y,F({\bf x}))$ where the function $F(\bf{ x})$ is calculated step wise as $F_m({\bf x})=F_{m-1}({\bf x})+\eta\gamma_m$ where $\eta$ is the learning rate, $\gamma_m$ is given by the loss function of the previous step.  

In usual ML, minimizing differences of predicted values and correct values are tried by minimizing the loss function\cite{Friedman01}. In our case, correct values correspond to the path with minimal action, whose solution is not known. Therefore, different from usual CNN, which was successful in pattern recognition etc. we need to do unsupervised trainings. Search of the optimum weight function of 20 FP actions is underway.

When the optimum weight function is obtained, CFT from $\bf u$ space to $\bf x$ space yields information of the anomalous scattering position of the phonon waves. Application to the medical research becomes in sight\cite{DSHSF22}. 

\newpage
When the mass gap of phonons remain, it becomes solitonic. The same will happen for neutrinos other than Majorana neutrinos. 
Solitonic wave propagations are made not only from charged Dirac spinors but also charge neutral Weyl spinors.  TR symmetric solitonic wave in $(2+1)D$ system shows explicit chiral symmetry violation.

\begin{acknowledgments}
S.F. thanks RCNP of Osaka University for the support of using super computer SQUID and Prof. M. Arai for the support of using a workstation in his laboratory. 
\end{acknowledgments}

\end{document}